\documentclass[preprint,superscriptaddress,amsmath,amssymb,12pt]{revtex4-1}

\makeatletter
\renewcommand{\p@subsection}{}
\renewcommand{\p@subsubsection}{}
\makeatother

\usepackage{graphicx}
\usepackage{dcolumn}
\usepackage{bm}
\usepackage{setspace}
\begin{document}
\newcommand{\angstrom}{\mbox{\normalfont\AA}}
\newcommand{\Arg}[1]{\mbox{Arg}\left[#1\right]}
\newcommand{\bb}{\mathbf}
\newcommand{\braopket}[3]{\left \langle #1\right| \hat #2 \left|#3 \right \
rangle}
\newcommand{\braket}[2]{\langle #1|#2\rangle}
\newcommand{\be}{\[}
\newcommand{\br}{\vspace{4mm}}
\newcommand{\bra}[1]{\langle #1|}
\newcommand{\braketbraket}[4]{\langle #1|#2\rangle\langle #3|#4\rangle}
\newcommand{\braop}[2]{\langle #1| \hat #2}
\newcommand{\dd}[1]{ \! \! \!  \mbox{d}#1\ }
\newcommand{\DD}[2]{\frac{\! \! \! \mbox d}{\mbox d #1}#2}
\renewcommand{\det}[1]{\mbox{det}\left(#1\right)}
\newcommand{\ee}{\]} 
\newcommand{\eg}{\textbf{\\  Example: \ \ \ }}
\newcommand{\Imag}[1]{\mbox{Im}\left(#1\right)}
\newcommand{\ket}[1]{|#1\rangle}
\newcommand{\ketbra}[2]{|#1\rangle \langle #2|}
\newcommand{\kp}{\arccos(\frac{\omega - \epsilon}{2t})}
\newcommand{\ldos}{\mbox{L.D.O.S.}}
\renewcommand{\log}[1]{\mbox{log}\left(#1\right)}
\newcommand{\Log}{\mbox{log}}
\newcommand{\Modsq}[1]{\left| #1\right|^2}
\newcommand{\nb}{\textbf{Note: \ \ \ }}
\newcommand{\op}[1]{\hat {#1}}
\newcommand{\opket}[2]{\hat #1 | #2 \rangle}
\newcommand{\occ}{\mbox{Occ. Num.}}
\newcommand{\Real}[1]{\mbox{Re}\left(#1\right)}
\newcommand{\sgn}{\text{sgn}} 
\newcommand{\so}{\Rightarrow}
\newcommand{\sol}{\textbf{Solution: \ \ \ }}
\newcommand{\thetafn}[1]{\  \! \theta \left(#1\right)}
\newcommand{\tin}{\int_{-\infty}^{+\infty}\! \! \!\!\!\!\!}
\newcommand{\Tr}[1]{\mbox{Tr}\left(#1\right)}
\newcommand{\kb}{k_B}
\newcommand{\rad}{\mbox{ rad}}
\title{Sublattice imbalance of substitutionally doped nitrogen in graphene}

\author{James A. Lawlor}
\email{jalawlor@tcd.ie}
\affiliation{School of Physics, Trinity College Dublin, Dublin 2, Ireland}

\author{Paul D. Gorman}
\affiliation{School of Physics, Trinity College Dublin, Dublin 2, Ireland}

\author{Stephen R. Power}
\affiliation{Center for Nanostructured Graphene (CNG), DTU Nanotech, Department 
of Micro-    
and Nanotechnology, Technical University of Denmark, DK-2800 Kongens Lyngby, 
Denmark}

\author{Claudionor G. Bezerra}
\affiliation{School of Physics, Trinity College Dublin, Dublin 2, Ireland}
\affiliation{Departamento de
Fisica, Universidade Federal do Rio Grande do Norte, Natal, RN, Brazil}

\author{Mauro S. Ferreira}
\affiliation{School of Physics, Trinity College Dublin, Dublin 2, Ireland}
\affiliation{CRANN, Trinity College Dublin, Dublin 2, Ireland}

\date{\today}

\begin{abstract}
%

Motivated by the recently observed sublattice asymmetry of substitutional 
nitrogen impurities in CVD grown graphene, we show, in a mathematically transparent manner,
that oscillations in the local density of states driven by the presence
 of substitutional impurities are responsible for breaking the sublattice 
symmetry. 
While these oscillations are normally averaged out in the case of randomly 
dispersed impurities, in graphene they have either the same, or very nearly the 
same, periodicity as the lattice. 
As a result, the total interaction energy of randomly distributed impurities 
embedded in the conduction-electron-filled medium does not vanish and is lowered 
when their configuration is sublattice-asymmetric. 
We also identify the presence of a critical concentration of nitrogen above which one should expect the sublattice 
asymmetry to disappear.
This feature is not particular to nitrogen dopants, but should be present in other impurities.
\end{abstract}
 
                 
\maketitle
\bibliographystyle{abbrv} 

\section{Introduction}
With remarkable physical properties, graphene is currently in the scientific 
limelight not only for possessing tremendous technological potential but also 
for having opened several avenues of basic science exploration \cite{geim_review,
 roadmap}. The honeycomb lattice structure of graphene and its two sublattices 
 are reponsible for a variety of novel physics phenomena.
 Recent theoretical reports suggest that doping one sublattice with vacancies \cite{vitor,vacanciesAsymm2}, adsorbates \cite{falko} and substitutional impurities 
 \cite{vitor, charlier-NL, bandGapAsymDoping, dopingAsymTransportProperties} 
might lead to the appearance of band gaps, and in the substitutional case 
induce spin-polarized current \cite{spinPolarized,spinPolarized2}. This unusual 
feature may pave the way to engineer the transport properties of doped graphene 
leading to a new species of transistor \cite{charlier-NL,transistors1,transistors2}.
Among the unresolved issues in the field is the recent observation in CVD 
grown graphene on Cu substrates that substitutional nitrogen dopants display a rather unusual 
asymmetry - heavily favouring sites on one of the two graphene sublattices \cite{zhaoScience,boron,terrones, zabet}.
Density Functional Theory (DFT) calculations by Zabet-Khosousi et al. \cite{zabet} suggest individual nitrogen impurities attaching to the edge of a 
graphene sheet prefer to occupy one sublattice over the other, which would lead to domains of segregation agreeing with the 
graphene grain boundaries. However, such a relationship between crystal grain and the sublattice segregation domains of the impurities
has not been seen experimentally.
 With this as a motivation, we address the problem of why substitutional impurities in graphene display
such an asymmetry in their configuration.
Besides offering a possible explanation for the observed sublattice asymmetry, our results suggest that this feature could be produced with 
other impurities under certain circumstances.  

Graphene sublattices, hereafter referred to as A and B, are triangular lattices that span the hexagonal structure when they are superimposed a suitable distance apart.
Being perfectly equivalent to one another, the only way that impurities may prefer one of the sublattices is if this sublattice symmetry is broken.
Symmetry breaking operations are known to induce some degree of segregation in the way impurities are spread across the structures.
Graphene edges, for example, can break the sublattice symmetry leading to a modulation in the spatial distribution of impurities \cite{segregation}.
In fact, theoretical calculations performed by some of the authors have suggested that it is possible to have one energetically preferential sublattice when impurities are near an edge.
While this might indicate a possible explanation for the puzzling asymmetry seen in the case of nitrogen doping,
such effects are strongly dependent on the edge geometry and only occur in the close vicinity of edges, which cannot explain 
the experimentally observed doping asymmetry across the entire sample.
Here we argue that, rather than edge effects, the substitutional impurities themselves are responsible for breaking the symmetry and for making one of the graphene
sublattices more energetically favourable than the other.
Similar symmetry breaking arguments have been put forward for the case of adsorbed impurities \cite{falko}, but the effect is less robust than the substitutional case.
Most importantly, in this manuscript we point to the existence of a critical impurity concentration beyond 
which the asymmetry disappears and one would expect to find dopants evenly distributed between sublattices.



\section{Model and Calculation Details}
To demonstrate that impurities are the key symmetry-breaking agents we must 
account for the total energy balance of a system with many impurities. 
We will start with the energy contribution from a single impurity, then a pair 
of impurities, and finally a system with a finite concentration of randomly distributed impurities.
To consider such a range of systems, a theoretical framework which can account 
for a scalable number of randomly placed impurities is essential. 
One such method can be found within the tight-binding (TB) model which allows 
us to describe the electronic structure of graphene as well as that of the 
substitutional impurities. 
The TB Hamiltonian for a pristine graphene sheet is ${\hat H} = \sum \vert j,\gamma
\rangle \, t \, \langle j^\prime, \gamma^\prime \vert$, where $t = -2.7 {\rm eV}$
is the electronic hopping between nearest neighbour carbon atoms, the 
basis $\vert j,\gamma \rangle$ represents a single atomic orbital located at a 
site that is identified by the unit cell $j$ in the sublattice $\gamma = { A} \,
 {\rm or} \, { B}$, and the sum runs over all nearest neighbours. 
One key quantity in this study is the total energy change that results from 
the introduction of a perturbing potential ${\hat V}$ due to the presence of 
substitutional impurities. 
The electronic contribution to the total energy change  ($\Delta {\cal E}$) of 
the system  is fully described by the Lloyd formula method \cite{LloydFormula} 
and results from the changes in the electronic density by the introduction of ${\hat V}$:
\begin{equation}
\Delta {\cal E} = {\frac{2}{ \pi}} \, \int_{-\infty}^{\infty} dE \, f(E) \, {\rm Im}
\ln \det {{\hat 1}
- {\hat {\cal G}} {\hat V}} \,\,,
\label{lloyd}
\end{equation}
where $f(E)$ is the Fermi function, ${\hat {\cal G}}$ is the single-particle 
Green Function (GF) associated with the TB Hamiltonian of a pristine graphene 
sheet and ${\hat 1}$ is the unit operator \cite{power&ferreira}.
The integrand above depends on the form of the perturbation ${\hat V}$, which for the 
case of a single impurity is ${\hat V} = \vert j_1, 
A \rangle \lambda \langle j_1, A \vert$, corresponding to a substitutional 
impurity with onsite energy $\lambda$ localized at the unit cell $j_1$ in 
sublattice $A$. 
Alternative forms of $\hat{V}$ that include extra matrix elements 
may be used but do not modify the results found here, due to the underlying 
symmetry contained in the perturbation potential. 
In the case of a single impurity, Eq.(\ref{lloyd}) becomes 
\begin{equation}
\Delta {\cal E}_1 = \frac{2}{\pi} \, \int dE \, {\rm Im} \ln ( 1 - {\cal
G}_{j_1,j_1}^{A,A} \, \lambda) \,\,,
\label{lloyd-1imp}
\end{equation}
where ${\cal G}_{j_1,j_1}^{A,A} = \langle j_1, A \vert {\hat {\cal G}} \vert j_1,
A \rangle$ is the diagonal matrix element of the single-particle GF, and we 
have omitted the integration range and the Fermi function for conciseness. 
Since the diagonal element of the single-particle GF carries no position 
dependence it is not necessary to solve the integral of Eq.(\ref{lloyd-1imp}) 
to show that the energy cost $\Delta {\cal E}_1$ is position independent. 

When a second impurity is added to the system, at some site $j_2$, on sublattice
 $\gamma_2$, the energy is not simply twice $\Delta {\cal E}_1$, but is given by
$\Delta {\cal E}_2 = 2 \, \Delta {\cal E}_1 + C_{1,2}^{\gamma_1, \gamma_2}$, 
where $C_{1,2}^{\gamma_1, \gamma_2}$ contains multiple scattering effects 
between the two impurities and is written as    
\begin{equation}
  C_{1,2}^{\gamma_1, \gamma_2} = \frac{2}{\pi} \int dE \, {\rm Im} \ln 
  \left({
  1 - 
\frac{\lambda^2 \,
{\cal G}_{1,2}^{\gamma_1,\gamma_2} \, {\cal G}_{2,1}^{\gamma_2,\gamma_1} }{ (1-{
\cal G}_{1,1}^{\gamma_1,\gamma_1} \lambda)
(1-{\cal G}_{2,2}^{\gamma_2,\gamma_2} \lambda)}  
}\right) \,\, . 
\label{C12}
\end{equation}
Since $C_{1,2}^{\gamma_1, \gamma_2}$ contains off-diagonal matrix elements of the GF, it will depend 
on the relative position of impurities 1 and 2, and we note that 
with only two impurities embedded in an infinite system the Fermi energy ($E_F$)
 will not be shifted from that of pristine graphene. 

\begin{figure}[t]
\centering
\includegraphics[width =0.7\textwidth]{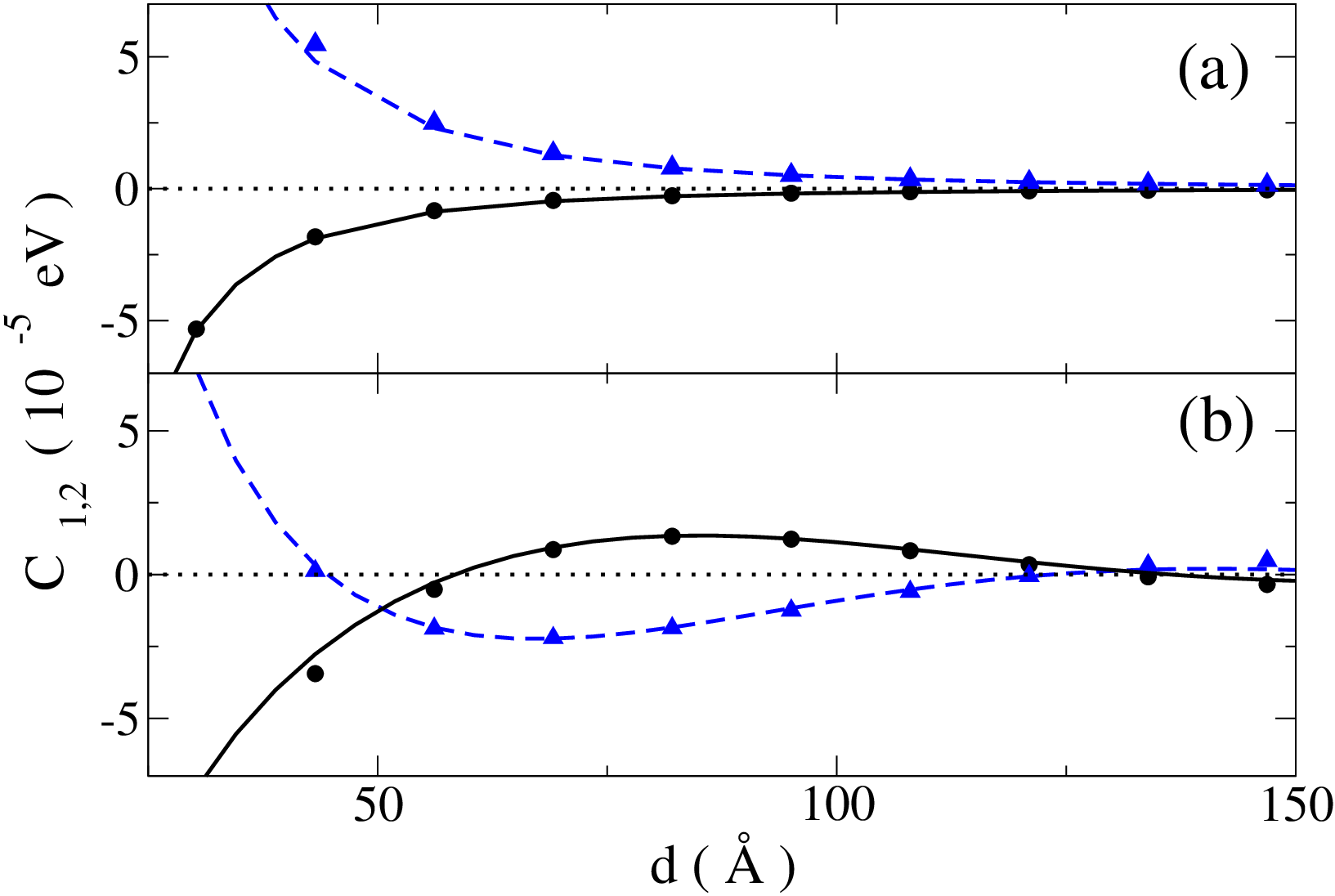}
\caption{$C_{1,2}^{\gamma_1, \gamma_2}$ plotted as a function of the separation between impurities,
(a) $E_F = 0$;  (b) $E_F = 0.135$eV, both using the impurity parameterisation $\lambda = -4 eV$. 
Solid (dashed) lines and circular (triangular) symbols represent $C_{1,2}^{A,A}$ 
($C_{1,2}^{A,B}$).
Symbols are the numerically evaluated values, whereas the lines follow the 
functional form $\alpha \cos{2 Q d + \phi} / d^2$}
\label{fig1}
\end{figure}

The functional form of $C_{1,2}^{\gamma_1, \gamma_2}$ has been derived based on the oscillations in 
the local density of states (LDOS) of graphene in the presence of a pair of 
substitutional impurities \cite{FO-james, power&ferreira}, and is seen to behave as 
a decaying sinuosoid $C_{1,2}^{\gamma_1, \gamma_2} = {\alpha} \cos\left({2 Q d + \phi}\right)/d^2$, 
where $\alpha$ , $Q$ and $\phi$ are functions of the Fermi Energy and $d$ is the separation
of the two impurities.
Fig.~\ref{fig1} plots $C_{1,2}^{\gamma_1, \gamma_2}$ against $d$ using the parameterisation $\lambda = -4 eV$, showing a negative value 
for $C_{1,2}^{A,A}$ and a positive value for $C_{1,2}^{A,B}$ at $E_F=0$ for all 
separations, i.e. it is energetically more favourable to accommodate a second 
impurity on the same sublattice at $E_F=0$, regardless of separation, while for $E_F\ne0$ the energetic 
favourability is separation dependent.
Note the lack of oscillations in Fig.~\ref{fig1}a despite its explicit dependence
on $\cos(2Qd)$. Here the oscillatory character of $C_{1,2}^{\gamma_1, \gamma_2}$ is suppressed by the 
commensurability of the possible separations and the oscillation period $Q = \cos^{-1}\left({ -\sqrt{1 - E_F^2} }\right)$.
Such a commensurability feature has been previously reported for nanotubes \cite{antc&ferreira} but is also prevalent in graphene \cite{crystals, power&ferreira}. 

\begin{figure}[t]
\centering
\includegraphics[width =0.7\textwidth]{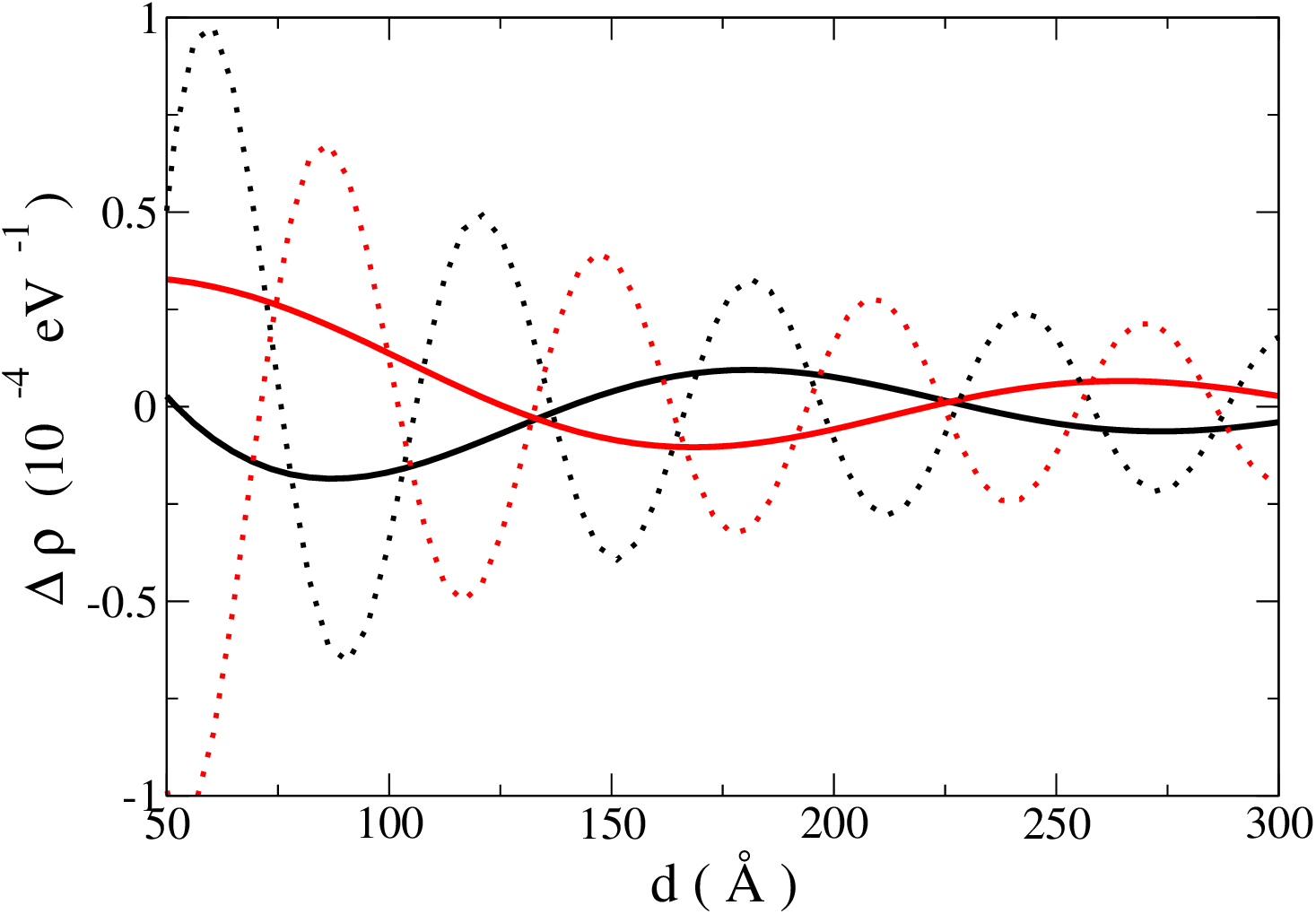}
\caption{Change in LDOS spectrum ($\Delta \rho$) at energies $0.1$eV (solid) and $0.3$eV (dashed) due to the presence of a single
nitrogen impurity on the black sublattice and parameterised by the on-site energy shift $\lambda = -4eV$,
as a function of distance from the impurity measured in the armchair direction for lattice sites
on the black (black data) and white (red data) sublattices. It can be seen here that the oscillations have a phase shift of approximately $\pi$ between each sublattice.}
\label{fig:ldos}
\end{figure}

The physical mechanism that drives this energetic preference for one of the 
sublattices can be understood as a result of the LDOS oscillations.
These oscillations arise from the presence of nitrogen dopants, which cause significant changes in the LDOS (See references \cite{lambin, terrones} and Fig. \ref{fig:ldos}).
In general terms, LDOS oscillations induce oscillations in 
other quantities (often referred to as Friedel Oscillations) that depend on the spatial distribution of impurities \cite{bacsi1,bacsi2,falkoFriedel}.
Because the total energy is one such a quantity, $C_{1,2}^{\gamma_1, \gamma_2}$ tends to oscillate 
between positive and negative values but the commensurability effect masks 
these oscillations in such a way that impurities on the same (different) 
sublattices always yield negative (positive) values.

Another way of understanding the energetic preference for one of the sublattices
 is to recall that graphene electrons are multiply scattered by the 
impurities. 
These interference effects between scatterers are responsible for minimizing 
the total energy.
In other words, impurities become aware of their mutual presence through an 
indirect interaction between them that is mediated by the conduction electrons 
of graphene. 
This is somewhat similar to the RKKY interaction that exists between localized 
magnetic moments embedded in a medium with conduction electrons. 
In fact, the theoretical framework presented here is analogous to the RKKY 
theory in graphene, the main difference being that there are no magnetic 
moments involved in this case.
For instance, it can be shown that when the integrand in Eq.(\ref{C12}) is 
expanded to lowest order in powers of $\lambda$ the expression for $C_{1,2}^{\gamma_1, \gamma_2}$ 
acquires exactly the same functional form as the RKKY coupling in 
graphene \cite{power&ferreira, crystals}.

\section{Sublattice Ordering at Finite Impurity Concentrations}
We now consider a dilute concentration of $N$ substitutional impurities 
randomly distributed in a graphene sheet.
The perturbation $\hat{V}$ will now consist of many additive terms of the form 
used to derive Eq.(\ref{lloyd-1imp}).
Following similar steps the total change in energy can be approximated as 
\begin{equation}
 \Delta {\cal E}_N \approx  N\Delta {\cal E}_1 + \sum_{i,j}C_{i,j}^{\gamma_i, \gamma_j},
\end{equation}
where the sum runs over all $N(N-1)/2 $ pairwise interactions in the 
system and terms beyond two-body interaction are neglected. 
It should be noted that despite the small magnitude of $C_{1,2}^{\gamma_1, \gamma_2}$, which is typically $10 ^{-6} eV$ to $10^{-4} eV$ at the average nearest neighbour separation,
the magnitude of the sum over all pairwise interactions, $\sum_{i,j} C_{i,j}^{\gamma_i , \gamma_j}$,
grows quadratically with the number of impurities in the system. This leads to a 
sizable correction to the energy balance for a large number of impurities 
 because of the commensurability effect discussed previously.
%

For the single and double impurity cases the Fermi Energy is not modified, however for low concentrations, $\rho$, the Fermi energy will shift as $E_F \sim \sqrt{\rho}$ and the pairwise impurity 
interaction will be oscillatory, resembling Fig.~\ref{fig1}b and Fig.~\ref{fig:ldos}. 
To demonstrate that asymmetric doping is indeed a feature of the multiple 
scattering effects between pairs of impurities, we shall consider the energy 
balance in two extreme limits - fully symmetric and fully asymmetric.
To do this we need to consider the concentration of impurities on each sublattice, A and B,
given by $\rho = \rho_A + \rho_B$.

Rather than considering a sum over all sites, we instead define a probability 
density as the total pairwise interaction energy with a single impurity, on site 1, from all 
other impurities. This density, which depends on the sublattice occupied by the impurity at site 1, 
can be calculated as
\begin{equation}
 \begin{split}
  E_A &= \sum_{j \neq 1} C_{1 j}^{A \gamma_j} = 2\pi \left({ \rho_A \int_{\sigma_{A,A}}^{\infty} r C_{1,2}^{A,A} dr +  \rho_B \int_{\sigma_{A,B}}^{\infty} r C_{1,2}^{A,B} dr   }\right),\\
  E_B &= \sum_{j \neq 1} C_{1 j}^{B \gamma_j} = 2\pi \left({ \rho_B \int_{\sigma_{B,B}}^{\infty} r C_{1,2}^{B,B} dr +  \rho_A \int_{\sigma_{B,A}}^{\infty} r C_{1,2}^{B,A} dr   }\right),
 \end{split}
\end{equation}
for impurities on sublattice $A$ and $B$ respectively, where $\sigma_{\gamma_1,\gamma_2}$ is the distance from a site on the $\gamma_1$ sublattice
to the nearest site on the $\gamma_2$ sublattice.
By symmetry we can write $\int_{\sigma_{A,A}}^{\infty} r C_{1,2}^{A,A} dr = \int_{\sigma_{B,B}}^{\infty}
r C_{1,2}^{B,B} dr = S_1(\lambda) $ and $\int_{\sigma_{A,B}}^{\infty} r C_{1,2}^{A,B}
dr =  \int_{\sigma_{B,A}}^{\infty} r C_{1,2}^{B,A} dr = S_2(\lambda)$.
The total pairwise interaction energy in the system, $\sum_{i,j} C_{i,j}^{\gamma_i , \gamma_j}$, can now be found by summing over all impurities
\begin{equation}
 \Delta E_n - n \Delta E_1 = \frac{1}{2} \left({ N_A E_A + N_B E_B   }\right),
\end{equation}
where $N_{A}= \pi R^2 \rho_{A}$ denotes the number of impurities on the $A$ 
sublattice (and similarly for $N_B$ and $B$) in a system with radius $R$, and the half prevents 
double-counting of interactions.
To find the impurity segregation that minimizes the system energy we let $\rho_A 
= f \rho$ and $\rho_B = (1-f) \rho$, where $f$ denotes the fraction of 
impurities occupying the $A$ sublattice.
We can now write the total pairwise interaction energy of the system per unit area as
\begin{equation}
 \mathcal{C} =  \frac{\sum_{i,j} C_{i,j}^{\gamma_i , \gamma_j}}{\pi R^2} = \pi \rho^2 ( 2 f^2( S_1- S_2) - 2f(S_1 - S_2) + S_1 ).
\label{eq:energytotal} 
\end{equation}

\begin{figure}[h]
\centering
 \includegraphics[width =0.6\textwidth]{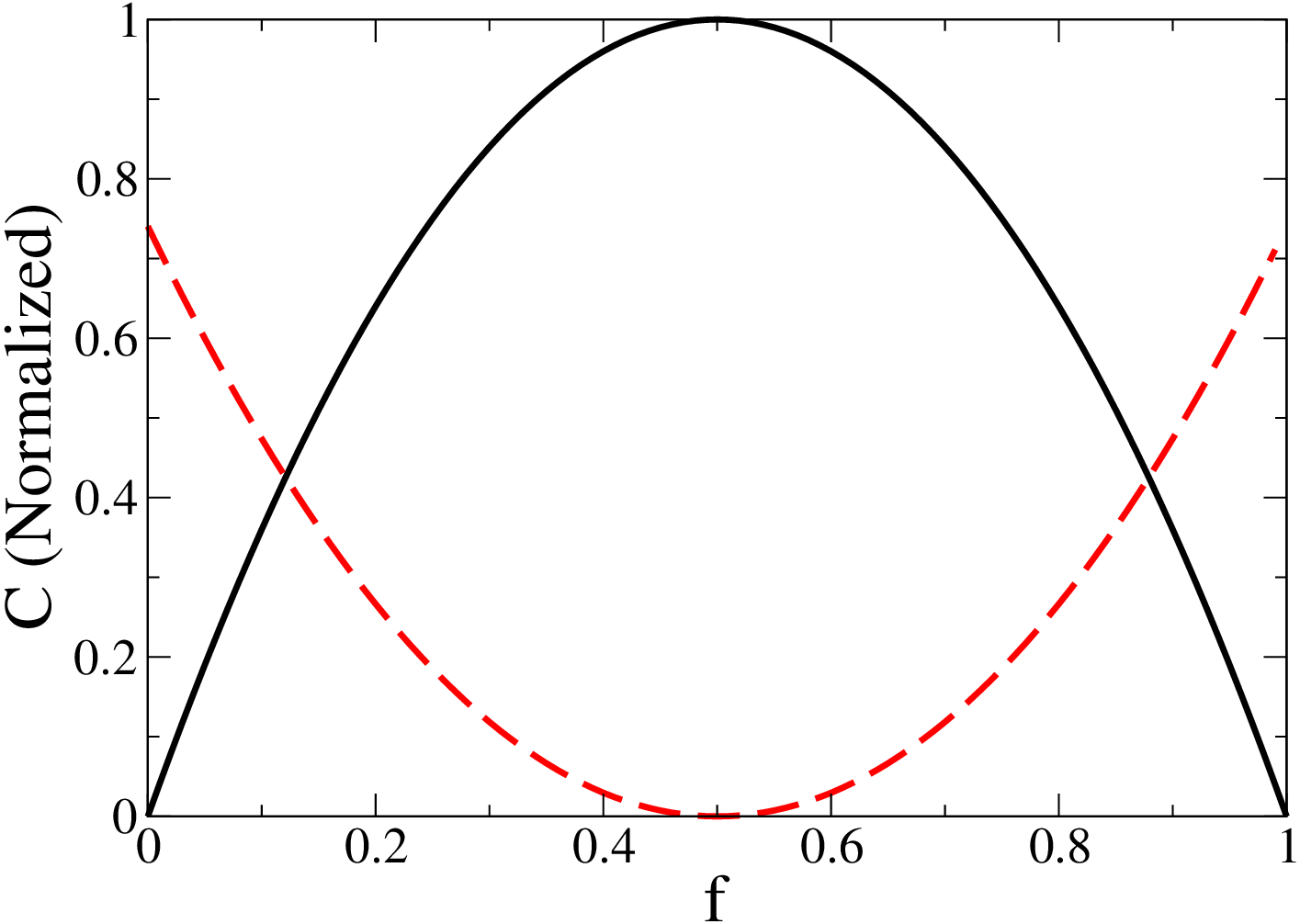}
\caption{A normalized plot of the total pairwise interaction energy per unit area, $\mathcal{C}$, against the 
fraction of impurities occupying sublattice $A$ with $\lambda= -4eV$ \cite{nitrogen_parameter}. The solid black (dashed red) line is calculated with a total 
concentration $\rho = 0.7 \%$ ($\rho = 0.9\%$), below (above) the critical concentration $\rho_c$. For concentrations above $\rho_c$ one would not expect to see sublattice segregation.}
\label{fig2}
\end{figure}

Provided $S_2(\lambda)>S_1(\lambda)$, this function has minima at $f = 0$ and $f = 1$ (Fig. \ref{fig2} solid black line)
meaning the most energetically favourable configuration is the fully asymmetric one.
This suggests the impurity sublattice segregation in graphene is indeed a 
feature of the multiple scattering effects between pairs of impurities, captured
 in the quantities $S_1(\lambda)$ and $S_2(\lambda)$. 
The dashed red line in Fig. \ref{fig2} shows the situation where $S_2(\lambda)<S_1(\lambda)$.
Here, the total pairwise interaction energy has a minimum at $f = \frac{1}{2}$ and the system does not favour any sublattice asymmetry.
 When $S_2(\lambda)=S_1(\lambda)$ a critical dopant concentration, $\rho_c$, is reached, beyond which one would
 not expect to see sublattice segregation. 
 


To illustrate the magnitude of $\mathcal{C}(f)$ we consider its value at some typical
  concentration, and also the difference between its maxima and minima and the size of this difference.
  As $\mathcal{C}$ is an energy per unit area, it is useful to consider the difference between the energies of the sublattice segregated ($f = 0$ or $f = 1$) 
  and sublattice unsegregated ($f = \frac{1}{2}$) states over the typical area occupied by a single impurity.
  For example, a concentration $\rho = 0.3\%$ and parameterisation $\lambda = -4eV$ yields a difference in energy of $-880 meV$.

\begin{figure}[h]
\centering
 \includegraphics[width =0.7\textwidth]{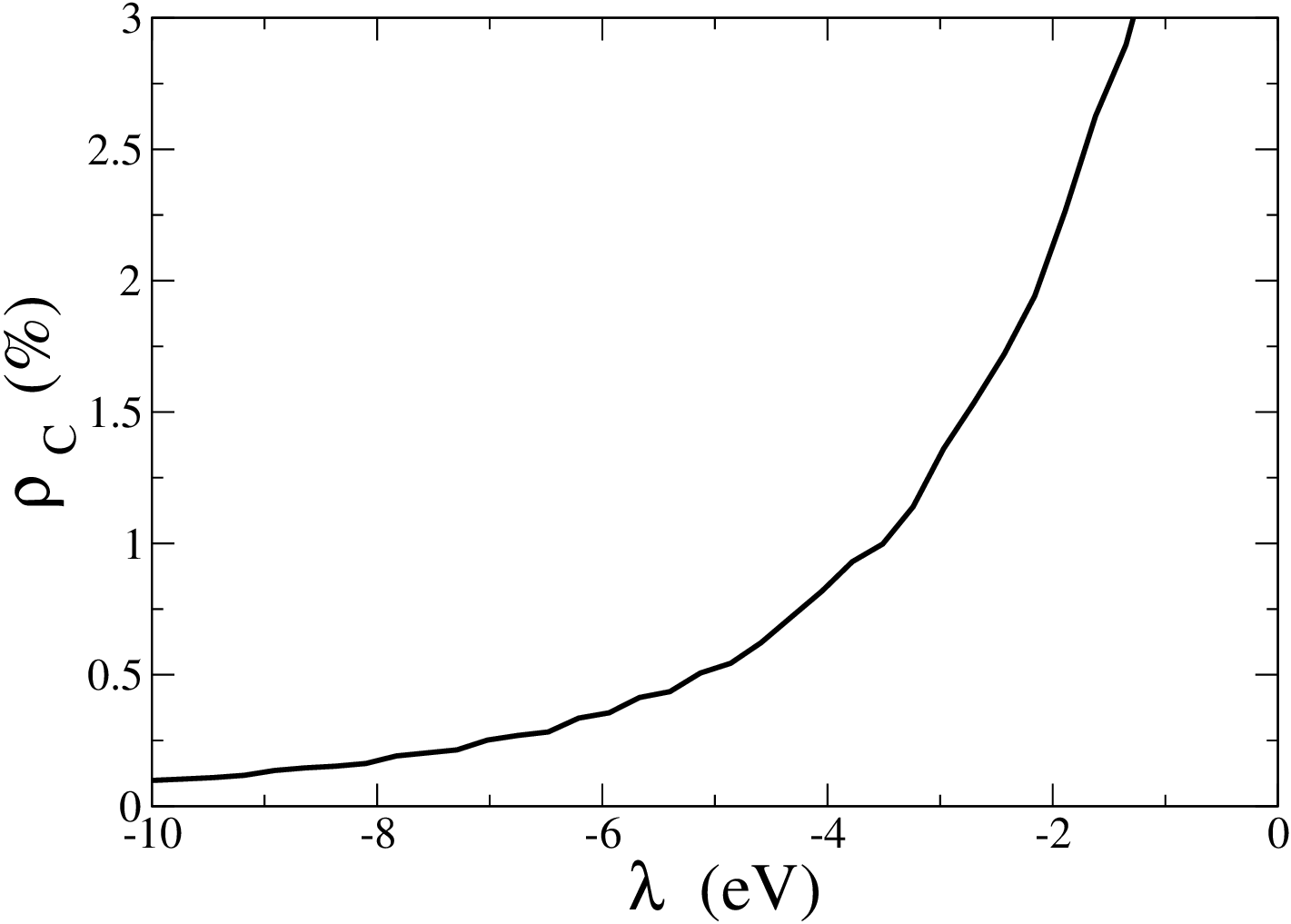}
\caption{Onsite energy $\lambda$ plotted against the critical concentration $\rho_c$.
 Nitrogen impurities are most commonly parameterised by $\lambda(N) = -4eV$ 
or $\lambda(N) = -10eV$, corresponding to critical concentrations of 0.8\% and 0.1\% respectively.}
\label{fig3}
\end{figure}

A numerical calculation of $S_1(\lambda)$ and $S_2(\lambda)$ as a function of $\lambda$ and $\rho$
gives us an insight into the types of impurities and concentrations
 that lead to sublattice segregation.
Fig.~\ref{fig3} shows a plot of critical concentration, $\rho_c$, versus onsite energy, $\lambda$.
 The onsite energy for nitrogen takes different values depending on the method of calculation.
The two most commonly used are $\lambda(N) = -4eV$, calculated through a self-consistent
tight binding model \cite{nitrogen_parameter}, and $\lambda(N) = -10eV$, through fitting of tight binding LDOS profiles to DFT results \cite{lambin}.
We see that for $\lambda(N) = -4eV$ the sublattice segregation should occur up until concentrations of $\sim 0.8\%$, as demonstrated in Fig.\ref{fig3},
well above the experimentally reported values of $\sim 0.1\% - 0.3 \%$ \cite{zhaoScience,boron,terrones,zabet}.
However, in the case of $\lambda(N) = -10eV$ the critical concentration falls to $0.1\%$, falling only slightly below 
the experimentally reported values.



Regarding the robustness of our calcuations, it is worth analysing how our results depend on the choice of the TB parameter $\lambda$ used to represent the nitrogen impurity.
According to Fig.\ref{fig3}, $\rho_c$ does depend on $\lambda$.
Therefore, we may conclude that the exact value of the critical concentration above which the sublattice asymmetry will no longer exist does depend on the choice of impurity parametrization.
However, Fig.\ref{fig3} also points to the existence of a critical concentration for all values of $\lambda$,
which indicates that the physical mechanism that we propose to explain the appearance of sublattice asymmetry in graphene also gives rise to a maximum concentration for which the effect is observable.
Consequently, our results point to a simple way in which the physical mechanism behind the sublattice asymmetry can be experimentally tested.
In other words, one may conclude that the electron mediated interaction between impurities is responsible for the observed sublattice asymmetry if such an effect disappears
when the nitrogen concentration is increased.
We emphasize that the existence of a critical concentration is robust to the values of the TB parameters
used to represent the electronic structure of graphene and of its substitutional dopants, however its value is indeed dependent on the parameterisation chosen.
Furthermore, the finite size of the experimental domains and their clearly defined edges are not handled in our model but can be explained by non-uniformities in the substrate structure, for example 
step edges, leading to a disruption in the inter-impurity interactions.

Our work here opens up the possibility of extending this method to other impurities, though this may prove more complicated.
Nitrogen, which sits next to carbon in the periodic table, is an especially simple dopant to work with in graphene, bonding similarly to carbon, and having only a single extra electron.
It would be natural to extend this to boron, which sits the other side of carbon, except that it is known to interact strongly with Cu substrates, destroying the effect \cite{boron}.
It is clear from Fig. \ref{fig3} that the critical concentration for nitrogen depends on the parameterisation $\lambda$,
which comes from the perturbing potential $\hat{V}$ that models the impurity. By extending this potential to include more terms e.g. hopping and nearest neighbour perturbations, 
we can model other impurities, where the exact values of the parameters can be calculated self-consistently \cite{nitrogen_parameter}.

\section{Conclusions}
In summary, we have shown that the LDOS oscillations that arise as a result of 
the presence of substitutional impurities are responsible for breaking the 
sublattice symmetry displayed by pristine graphene.
When considered over a large range of distances they are averaged out and are 
often disregarded for cases of randomly dispersed impurities. 
However, in graphene such oscillations have either the same periodicity of the 
lattice or are very close to it, leading to a net result that affects the 
behaviour of several physical quantities in doped graphene.
One such quantity is the indirect interaction energy that exists between 
impurities embedded in a conduction-electron-filled medium. In graphene this indirect interaction yields a 
finite contribution, which in turn drives the doping asymmetry.
Our model predicts the existence of a critical concentration for nitrogen dopants beyond which the sublattice asymmetry should no longer be visible.
While the exact value for such a concentration is not accurately predicted by our simple model, it nevertheless suggests that concentrations above 1\% are already above this critical limit.
Most importantly, from a qualitative point of view,
the mechanism we propose to explain the sublattice asymmetry of substitutional dopants in graphene leads to a simple consequence that can be easily tested experimentally.
Should the asymmetry cease to exist as dopant concentrations are increased,
we may conclude that the doping asymmetry arises due to fluctuation of the LDOS that break the natural symmetry that otherwise exists between the two sublattices of graphene.


\section{Acknowledgements}
The authors acknowledge financial support from the Programme for Research in 
Third Level Institutions (PRTLI), Science Foundation Ireland (Grant No. SFI 11/
RFP.1/MTR/3083), the Center for Nanostructured Graphene (CNG) which is sponsored
 by the Danish National Research Foundation, Project No. DNRF58, and CAPES (
Grant No. 10144-12-9), FAPERN and INCT of Space Studies.

\end{document}